%% file: imada.tex
\documentstyle[aps,twocolumn]{revtex}
\catcode`\@=11

\def\eqlt{\mathrel{\mathpalette\@vereq<}}  
\def\eqgt{\mathrel{\mathpalette\@vereq>}}  
\def\@vereq#1#2{\lower2.5pt\vbox{\baselineskip0pt \lineskip-.5pt
 \ialign{$\m@th#1\hfil##\hfil$\crcr#2\crcr{=}\crcr}}}

\newcommand{\simle}{\ \raise.3ex\hbox{$<$}\kern-0.8em\lower.7ex\hbox{$\sim$}\ }
\newcommand{\simge}{\ \raise.3ex\hbox{$>$}\kern-0.8em\lower.7ex\hbox{$\sim$}\ }

\begin{document}
\draft
\twocolumn[\hsize\textwidth\columnwidth\hsize\csname @twocolumnfalse\endcsname
\title{Superconductivity from Flat Dispersion Designed in Doped Mott Insulators }
\author{Masatoshi Imada$^1$ and Masanori Kohno$^2$}
\address{${}^1$Institute for Solid State Physics, University of Tokyo, 
Roppongi, Minato-ku, Tokyo 106-8666, Japan}
\address{${}^2$ Mitsubishi Research Institute, Inc., Ootemachi, Chiyoda-ku, Tokyo, 
100-8141, Japan}
\date{\today}
\maketitle

\begin{abstract}
Routes to enhance superconducting instability are explored for doped Mott 
insulators.  With the help of insights for  criticalities of metal-insulator 
transitions, geometrical design of lattice structure is proposed to control the 
instability.   A guideline is to explicitly make flat band dispersions near the 
Fermi level without suppressing two-particle channels.  In a one-dimensional 
model, numerical studies show that our prescription with finite-ranged hoppings 
realizes large enhancement of spin-gap and pairing dominant regions.  We also 
propose several multi-band systems, where the pairing is driven by intersite 
Coulomb repulsion.
\end{abstract}

]
{
Basic properties of doped Mott insulators have been a subject of recent 
continued studies~\cite{RMP}.  One of the  goals of the studies is to find ways 
to design instabilities such as magnetism and superconductivity by controlling 
material parameters in a realizable way.   It is desired to control the 
instabilities by utilizing inherent character of the doped Mott insulators and 
the critical nature of metal-to-Mott insulator transitions.  In this letter, 
possible prescriptions to control the instabilities are proposed.  

For single-band Hubbard and $t$-$J$ models on a square lattice, simple scaling 
properties have been observed numerically near
the transition from metals to Mott
insulators~\cite{Imada1995,Imada1999,RMP}, implying nontrivial and
singularly momentum-dependent correlation effects, where
single-particle excitations around some particular points as $(\pi,0)$
and $(0,\pi)$ in the momentum space play crucial roles with the
emergence of a flat dispersion~\cite{Assaad99}.  This criticality
extends to 20-30\% doping range while appears only below some fraction of the 
scale of the exchange interaction $J$ which is also the scale of the effective 
bandwidth.  The flat dispersion around ($\pi$,0) appears at similar scales.   
The flat dispersion was also numerically observed in the 2D Hubbard model in 
earlier works by Dagotto et al~\cite{Dagotto}, Bulut et al.~\cite{Bulut} and 
Preuss et al.~\cite{Preuss} with comparison to universally observed flat 
dispersions in angle-resolved photoemission data of the high-Tc 
cuprates~\cite{Gofron,Shen}.  
Although the observed flat dispersion is certainly resulted from a strong 
correlation effect with strong damping~\cite{Imada1999,Dagotto,Bulut,Preuss}, 
the microscopic mechanism for the flat dispersion still waits for more complete 
understanding.   
Furthermore, its persistence in multi-band or more complex systems has not been 
well studied.  

In this letter we discuss that promotion of the above scaling behavior and the 
flat dispersion offers a way to control potential instabilities. We show that 
even when a flat {\it band} dispersion is designed near the Fermi level by 
controlling lattice geometry and parameters, it enlarges the critical region 
under the suppression of single-particle coherence in the
proximity of the Mott insulator mentioned above, thereby enhances the 
instability.   
One might argue that if a flattened dispersion is designed, it simply
makes the correlation effects relatively larger only through the
change in the ratio to the effective bandwidth with the enhanced density of 
states.  This is, however,
not the whole story on the verge of itinerant and
correlation-induced localized states.  The metallic excitations are determined 
from the coherent one near the Fermi level while the two-particle processes 
including the superexchange interaction $J$ is rather determined mainly from a 
local, incoherent origin in the real space when the electron correlation is 
strong.   It opens a possibility of enhancing the two-particle instability by 
suppressing the dispersion only near the Fermi level simultaneously keeping the 
amplitude of the two-particle processes large.  By the suppression of only the 
single-particle coherence, two-particle processes work selectively and 
effectively since competition processes such as pair-breakings and damping by 
the single-particle channel are suppressed.  We find no particular dependence on 
dimensionality for this mechanism.  This is, however, not possible in 
single-band models with nearest-neighbor hopping because the two-particle 
processes is not independent of the band dispersion near the 
Fermi level.  For example, $J$ and the so-called three-site terms in the $t$-$J$ 
models are simply scaled by the transfer $t$ as $t^2/U$ in the strong 
correlation expansion.  We discuss below how the independent control can be made 
in more complex systems.

The flattened dispersion makes degenerate excitations and may cause various 
instabilities.  In this
context, we note that ferromagnetic instabilities by
the flat band has been extensively studied~\cite{Lieb,Mielke,Tasaki}.
However, we consider here only the cases with the singlet
ground state at half filling, where the ferromagnetic instability is suppressed.

In 1D $t$-$J$ models, the phase diagram in the parameter space of $J$/$t$ and 
electron concentration $n$ shows a general tendency of stronger pairing 
instability namely the larger Tomonaga-Luttinger exponent $K_{\rho}$ with spin 
gapped excitation for larger $J$/$t$, though the phase separation interrupts the 
enhancement~\cite{Ogata91,Ammon95,Nakamura97}.  Note that the pairing 
correlation is the most dominant if $K_{\rho} > 1$. The general tendency for 
enhanced pairing instability for large $J/t$ also holds in 2D~\cite{Dagotto}. 
These are consistent with the mechanism we discussed above.  We define the 
ordinary $t$-$J$ model with the three-site terms:
\begin{eqnarray}
{\cal H} & = & - t \sum_{\langle ij\rangle} (\tilde 
{c}^{\dagger}_{i\sigma}\tilde{c}_{j\sigma}+ {\rm H.c.})  + J \sum_{\langle 
ij\rangle}({\bf S}_i\cdot{\bf S}_j -\frac{1}{4}n_in_j)\nonumber \\
& & - \frac{J}{4}\sum_{\langle 
ijk\rangle}(\tilde{c}^{\dagger}_{i\sigma}n_{j-\sigma}\tilde{c}_{k\sigma}-\tilde{
c}^{\dagger}_{i\sigma}\tilde{c}^{\dagger}_{j-\sigma}\tilde{c}_{j\sigma}\tilde{c}
_{k-\sigma} + {\rm H.c.}) \label{eq:H}
\end{eqnarray}
where $\langle ij\rangle$ and $\langle ijk\rangle$ are nearest neighbors and 
$\tilde{c}^{\dagger}_{i\sigma}(\tilde{c}_{i\sigma})$ creates(annihilates) 
electrons with constraint to exclude the double occupancy.  
If the band dispersion is suppressed only near the Fermi level, the effective 
transfer decreases while $J$ can be retained mediated by incoherent but 
dispersive part.  This is indeed possible by introducing longer-ranged transfer. 
 We introduce the 1D model with the third, fifth and seventh neighbor transfers,
$t_3, t_5$ and $t_7$, respectively and for the exchange part, $J_i =
J({t_i}/{t})^2$ accordingly, in addition to  $t$ and
$J$~\cite{Kohno99}. We design flat noninteracting dispersion
$\epsilon(q)$ by optimizing  $t_3, t_5$ and $t_7$ to make terms up to
the sixth order vanish in the wavenumber $q$ around $q=\pm \pi/2$, the
Fermi level at half filling.  The spin-gap boundary and the exponent $K_{\rho}= 
\sqrt{\pi
  Dn^2\kappa/4}$ were calculated by exact
diagonalization of the Hamiltonian up to 16 sites, where the Drude
weight $D$ and the compressibility $\kappa$ were calculated at filling
$n$ following the procedure in the
literature~\cite{Ogata91,Ammon95,Nakamura97}.  For the spin
gap boundary, we used the level crossing method for accurate
estimates~\cite{Nakamura97}.  The results have practically no system size 
dependence implying a reliable estimate of the thermodynamic results.  They show 
remarkable enhancement
of both the pairing correlations and spin gap region as in
Fig.~\ref{Fig1}, where the phase separation is absent.  It shows the mechanism 
we proposed is indeed effective. We have also examined other types of band 
modifications and confirmed that the above mechanism mainly determines the 
enhancements~\cite{Kohno99}.  
We have found that the corresponding Hubbard-type model with the same
distant-ranged transfers also show a similar enhancement.  However, the system 
sizes we could study is not large enough for reliable estimates of 
the thermodynamic limit.  We note that the enhancement of pairing was also 
reported in the ladder model~\cite{Yamaji,Kuroki,Scalapino}, when the Fermi 
level lies near the top(bottom) of bands, where a flattening is present.  
Our finding suggests one prescription, namely, look for a tuned flatter 
dispersion arranged only near the Fermi level of the doped Mott insulator in 
quasi one-dimensional conductors. Since the mechanism itself is not confined to 
quasi-one dimensionality, it opens various possibilities with rich physics along 
this line. 

Next, we analyze several multi-band models with nontrivial band-flattening 
effects to get further insight on this issue and to provide hints for material 
synthesis.
We introduce an extended  Hubbard model 
\begin{eqnarray}
{\cal H}_H & = & -t\sum_{\langle 
ij\rangle}(c^{\dagger}_{i\sigma}c_{j\sigma}+H.c.) -t'\sum_{\langle 
lm\rangle}(c^{\dagger}_{l\sigma}c_{m\sigma}+H.c.) \nonumber \\
 & + &U \sum_i (n_{i\uparrow}-\frac{1}{2})(n_{i\downarrow}-\frac{1}{2}) + 
\sum_{ij}V_{ij} n_in_j \label{eq:2.b.2.1c} 
\end{eqnarray}
where $c^{\dagger}_{i\sigma}(c_{i\sigma})$ represents the creation 
(annihilation) operator of an electron at site 
$i$ with spin $\sigma$  and $n_{i\sigma} \equiv
c^{\dagger}_{i\sigma}c_{i\sigma}$ and $n_i  \equiv \sum_{\sigma}
n_{i\sigma}$ are number operators.  
The lattices are defined below in Figs.~\ref{Fig2a} -~\ref{Fig2d} where $t$ 
connects bonds with solid lines and $t'$ connects broken bonds.  

The first example belongs to a category of one-quarter depleted square lattice.  
Our model is described by the Hamiltonian (2) with the lattice
structure illustrated in Fig.~\ref{Fig2a}.   When
$t'=0$, it reduces to a lattice considered by Lieb~\cite{Lieb} if isolated spins 
on a quarter of lattice points are depleted.  It has ferromagnetic ground state 
at half filling as Lieb
proved.  For $t'\neq 0$, however, the ground state becomes
singlet at half filling due to the Lieb-Mattis
theorem~\cite{LiebMattis}.  For smaller $t'/t$, the noninteracting dispersion 
shows flattening in
the middle two bands among four in total.  The two flattened dispersions are 
given by 
$E  =  \pm\sqrt{-\frac{1}{2}(\Xi - 
\sqrt{(\Xi+2\sqrt{\Gamma})(\Xi-2\sqrt{\Gamma})})}$, 
with $\Xi=(v^2+w^2-u^2)\cos{(k_x-k_y)}-2u^2(\cos{k_x}+\cos{k_y})-3u^2-v^2-w^2,
\Gamma = w^2u^2(\cos{k_x}-\cos{k_y})^2,
w=2tt'/\sqrt{t^2+t'^2},
u=\sqrt{t^2+t'^2},
v=({t^2-t'^2})/{\sqrt{t^2+t'^2}}.$
In the limit of $t' \rightarrow 0$, these two bands become completely flat.
However flat the band becomes with small $t'/t$, the superexchange interaction 
on the solid bonds in Fig.~\ref{Fig2a} is unchanged and favors the 
antiferromagnetic 
order at half filling.  This lattice structure may 
favor  triplet pairing instability near $t'=0$ in contrast to the cases below, 
due to spin polarization of the flat nonbonding band.

The second model has slightly different lattice from the case above as 
illustrated in Fig.~\ref{Fig2b}.   
In this bipartite structure, the number of connected $A$ and $B$
sublattice points are equal even at $t'=0$ after
removing isolated spins, where the ground state is singlet at half filling and 
the Mott insulating state with the antiferromagnetic order is expected in the 
thermodynamic limit. The flattening of the band is not complete in this case, 
but still has an extended region of flat plateau.

The third example is illustrated in Fig.~\ref{Fig2c}.  The noninteracting
bands consist of a dispersive bonding and a completely flat
antibonding bands. Under
electron doping, carriers go into the flat band.   Because of a strong 
frustration in contrast to the other cases, the
magnetic correlation may be suppressed.

For the fourth system, the lattice structure is illustrated in Fig.~\ref{Fig2d}. 
 The square lattice structure in Fig.~\ref{Fig2d} is not important and the 
following argument applies to any other bipartite lattices  if the 
3-site unit cell has the same structure.   The band structure consists of 
antibonding, nonbonding and bonding bands from high to low energies given by 
$
\varepsilon_a  =  -tC_1 + \sqrt{(tC_1)^2+2t'^2}, \label{2a}
\varepsilon_n  =  0, \label{2b}
\varepsilon_b  =  -tC_1-\sqrt{(tC_1)^2+2t'^2}, \label{2c}
$
respectively with $C_1=\cos k_x + \cos k_y$ and an energy gap 
$\Delta_g  \equiv  -2t+\sqrt{4t^2+2t'^2}$.
The noninteracting ground state at half filling is given by filled bonding and 
half-filled 
nonbonding bands.  

Below we discuss a general aspect of interaction effects more or less
valid in all the above cases although we first take the fourth system as an 
example.  For nonzero $U$,  the flat band splits into upper and lower Hubbard 
bands.   At half filling, the lower Hubbard band of this nonbonding band is 
filled leading to  the Mott insulator.  Each cell is occupied by precisely one 
electron in the nonbonding orbital and an exchange interaction between these 
nonbonding electrons in the higher order in $U$ may stabilize the 
antiferromagnetic order.   In the perturbation expansion in $U$, we have mixing 
of antibonding  components into the filled bonding band.   However, in the 
perturbation, the number of nonbonding electrons at each cell can be changed 
only by two so that in all the cells, precisely one nonbonding electron is kept 
up to the infinite order in $U$.  Even away from half filling, when we once 
assign configuration for singly occupied and empty sites for nonbonding 
orbitals, they do not have dynamics, where macroscopic degeneracy remains.  If 
the perturbation in $U$ converges, the system remains insulating up to the 
doping concentration  $\delta=1/3$.  Intracell but intersite interaction plays a 
similar role to $U$.  

An important process to kill the insulating state arises from
the intercell Coulomb repulsion $V_{|i-j|}$ ($i\not= j$) between the
$i$-th and $j$-th cells.  For simplicity, we take $V$ between
the same sublattice points 1, 2, or 3 in Fig.~\ref{Fig2d}.  Other combinations 
play a similar role.  
 Away from half filling, there appear empty and doubly occupied cells in the 
nonbonding orbitals.  In the first order perturbation in $V_{|i-j|}$, a pair of 
electrons (holes) on the singly-occupied nonbonding orbitals at sites $i$ and 
$j$ each is excited to the itinerant antibonding (bonding) band, where the 
excitation energy is roughly $2\Delta_g$.   In the second order in $V$, a pair 
hopping from $i,j$ cells to $l,m$ cells occurs if $l$ and $m$ cells have 
initially no nonbonding electrons. 
Here we show the perturbation expansion up to the second order for the third 
model (Fig.~\ref{Fig2c}):  
\begin{eqnarray}
{\cal H}_{\rm eff} & = & 
\frac{1}{2}\sum_{i\delta}V_{\delta}n_{2,i}n_{2,i+\delta} \nonumber \\
& &-\sum_{\stackrel{\scriptstyle i\delta}{l\delta'}}
V_{\delta}V_{\delta'}f_{il\delta\delta'}a_{2,l+\delta',\sigma'}^{\dagger}a^{\dag
ger}_{2,l,\sigma}a_{2,i,\sigma}a_{2,i+\delta,\sigma'},\nonumber
\end{eqnarray}
\[
f_{il\delta\delta'} = 
\sum_{kk'}\frac{\cos[k'(r_{l+\delta'}-r_{i+\delta})]\cos[k(r_l-r_i)]}
{2t'-4t(\cos k + \cos k')}
\]
where $a^{\dagger}_{2,i,\sigma}$ creates a flatband electron at site $i$ with 
spin $\sigma$.  Similar effective Hamiltonians are derived for the other models.  

 From the second order process, the localized nonbonding electrons
 melt.  It also induces superexchange type magnetic interaction from
 the charge diagonal part.  
When $V$ has some extended range, the superconducting order with the pairing 
form factor with this range may occur.  If the range becomes longer, the 
magnetic interaction becomes more frustrated and the magnetic correlation is     
 suppressed.  The longer-ranged Coulomb interaction also helps to inactivate the 
first order process in $V$.  This is understood from the limit of infinite-range 
$V$.  If $V_{|i-j|}$ does not depend on the combination $(i,j)$, the first order 
process just gives a constant independently of the electron configuration.  If 
$V_{|i-j|}$ is given by a screened form $\sim [{\rm exp}(-r_{ij}/R)/r_{ij}]V_0$ 
with the range $R$, the second order process generates averaged kinetic energy 
in the order $(V_0^2/\Delta_g)\delta^2(R/a)^2(t/\Delta_g)^2$ in 2D where $a$ is 
the lattice constant.  The average Coulomb repulsion energy generated from the 
first order process is roughly $V_0/\delta^{1/d}$ in $d$ dimensions.   Then the 
pairing interaction in the second order dominates for larger $R$.  
By a rough estimate inferred from the melting transition of the Wigner 
crystal~\cite{Imada1984}, the dominance may well be realized for $R/a\simge 3$ 
at $\delta \sim 0.1$ and $V_0/\Delta_g \sim \Delta_g/t \simle 1$.   
If this condition is not well satisfied, the
charge ordering may compete with the superconductivity. 
A related multi-band pairing mechanism was pointed out 
before~\cite{Kondo,Yamaji}, where the pair transfer through $U$ drives the 
s-wave superconductivity. Although several other proposals for pairing in 
multiband models with $V$ have been made~\cite{Varma,Stechel}, they are rather 
different from ours in the sense that our mechanisms come from (interband) pair 
transfers from flattened band near the Fermi level to dispersive channel.
An artificial but helpful limit to understand the instability in our mechanism 
is the case of infinite-range transfer with the uniform amplitude $t/\sqrt{N}$ 
for system size $N$.  
In this case, the mean field  solution  with the order parameter $\Delta = 
\sum_k f(k)c_{k\sigma} c_{-k\sigma'}$ with 
$f(k)=\sum_{ij}V_{|i-j|}e^{ik|i-j|}/N$
becomes the correct description of the superconducting ground state, where the 
Hamiltonian has the form
${\cal H} = -\frac{1}{N}\Delta^{\dagger}\Delta\frac{t^2}{t'^3}.$
This gives an example of superconducting order derived from purely
repulsive interaction. 
Up to this discussion, the triplet and singlet pairings are degenerate. This is 
lifted by perturbation such as the frustrated magnetic interaction.  
 
It is tempting to say that the above designed flattening seems to be an extreme 
limit of the high-$T_c$ cuprates.  In our models, the flat dispersion is 
separated  from the dispersive bands.  The high-$T_c$ cuprates seem to be more 
similar to the fourth model at large $U$, where the dispersive (anti)bonding and 
the flat nonbonding bands coexist at the Fermi level.

It would be desired to seek for materials which follow the prescription
presented here.  For the first and second systems, regular 1/4
depletion of square lattice or substitution (for example one quarter
of Cu ions with nonmagnetic ions  on the CuO$_2$ lattice) may
satisfy the requirement.  It is interesting that this structure has
some similarity to the charge ordered  (or stripe) phase suggested in
La-based high-$T_c$ cuprates~\cite{Tran1995}. This connection provides a new
view on the stripe problem~\cite{Imada99}.  The fourth model may be realized by 
3-layered structure with top and bottom layers designed to remove ligand atoms 
such as oxygen or substituted with different ligand atoms to suppress transfer 
through them in those planes.

One might argue that, with flat dispersions, impurity or phonon effects were 
serious in real materials.  However, when the coherent single-particle 
excitations are suppressed, the localization effect is rather determined under 
the competition to the two-particle process.  The localization effect is reduced 
if the two-particle hopping process is retained larger.  Anyhow impurity and 
phonon effects near the Mott insulator in the present situation are important 
open problems left for future studies.

Our proposal to enhance the pairing instability is summarized:
Design flat band near the Mott insulator by retaining the preferred
two-particle process. Due to the enhanced degeneracy and the suppression of 
single-particle process, the doped system gets
stronger instability in the two-particle process. We have numerically shown that 
this mechanism and prescription work in an example of 1D models with 
longer-ranged transfers.  We have proposed several multi-band lattices, where 
the intersite Coulomb
repulsion generates such two-particle processes through interband pair
hoppings.  Because the coherent single-particle process is absent,
small intersite interaction immediately corresponds to
the strong coupling limit of the pairing where the Fermi liquid state does not 
exist.

\section*{Acknowledgements}
The authors thank H. Asakawa and H. Tsunetsugu for useful discussions.  This 
work is supported by a Grant-in-Aid for ``Research for the Future" Program from 
the Japan Society for the Promotion of Science under the project 
JSPS-RFTF97P01103.  

\input psbox.tex
\begin{figure}
$$ \psboxscaled{500}{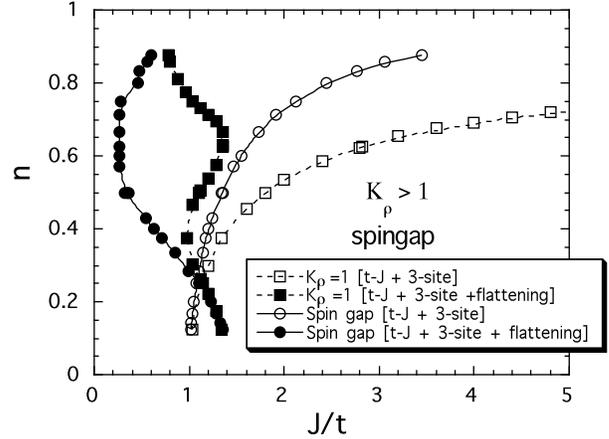} $$
\caption{Phase boundary of the spin gap region (circles) and the contour lines 
(with squares) for the Tomonaga-Luttinger exponent $K_{\rho} =1$.  The case for 
the $t$-$J$ model with the 3-site term (open symbols) and that of the same model 
but with optimized dispersions by $t_3=0.6, t_5=0.2$ and $t_7=1/35$ and 
resultant $J_3$,$J_5$ and $J_7$, where $J_i = Jt_i^2$ (filled symbols).}
\label{Fig1}
\end{figure}

\begin{minipage}[t]{4cm}
\begin{figure}
$$ \psboxscaled{300}{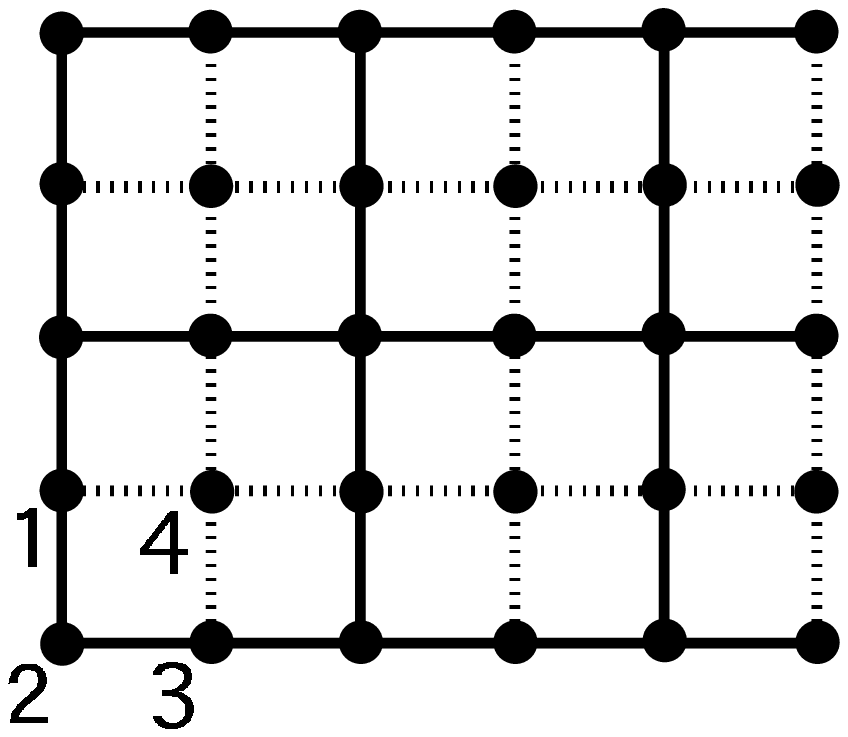} $$
\caption{(a)The first model with regular 1/4 depleted structure. }
\label{Fig2a}
\end{figure}
\end{minipage}
\begin{minipage}[t]{4cm}
\begin{figure}[hbt]
$$ \psboxscaled{300}{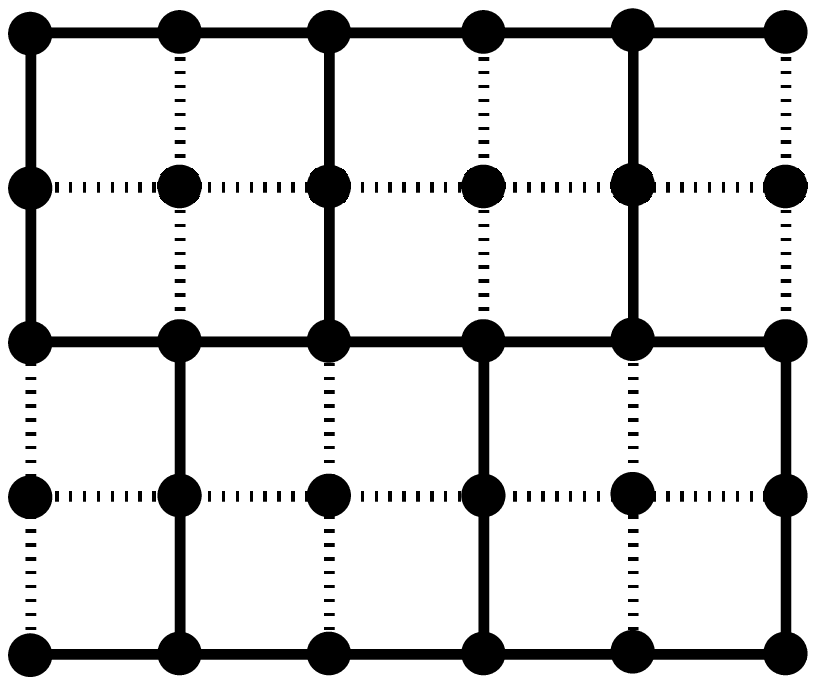} $$
\caption{The second model with alternating 1/4 depleted structure.}
\label{Fig2b}
\end{figure}
\end{minipage}

\vspace{-5mm}
\begin{figure}
$$ \psboxscaled{500}{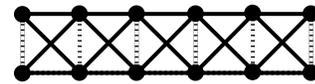} $$
\caption{The third model, ladder structure with diagonal transfer.}
\label{Fig2c}
\end{figure}

\begin{figure}
$$ \psboxscaled{450}{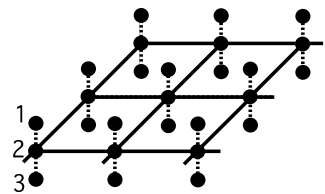} $$
\caption{The fourth model, decorated square lattice.}
\label{Fig2d}
\end{figure}
}
\end{document}

%% file: psbox.tex
\def\temp{1.35}%
\let\tempp=\relax
\expandafter\ifx\csname psboxversion\endcsname\relax
  \message{PSBOX(\temp)}%
\else
    \ifdim\temp cm>\psboxversion cm
      \message{PSBOX(\temp)}%
    \else
      \message{PSBOX(\psboxversion) is already loaded: I won't load
        PSBOX(\temp)!}%
      \let\temp=\psboxversion
      \let\tempp= 
    \fi
\fi
\tempp
\message{by Jean Orloff: loading ...}
\let\psboxversion=\temp
\catcode`\@=11
%
%
\def\psfortextures{
\def\PSspeci@l##1##2{%
\special{illustration ##1\space scaled ##2}%
}}%
\def\psfordvitops{
\def\PSspeci@l##1##2{%
\special{dvitops: import ##1\space \the\drawingwd \the\drawinght}%
}}%
\def\psfordvips{
\def\PSspeci@l##1##2{%
\d@my=0.1bp \d@mx=\drawingwd \divide\d@mx by\d@my
\includegraphics{##1\space}}}%
\def\psforoztex{
\def\PSspeci@l##1##2{%
\special{##1 \space
      ##2 1000 div dup scale
      \number-\psllx\space\space \number-\pslly\space\space translate
}}}%
\def\psfordvitps{
\def\dvitpsLiter@ldim##1{\dimen0=##1\relax
\special{dvitps: Literal "\number\dimen0\space"}}%
\def\PSspeci@l##1##2{%
\at(0bp;\drawinght){%
\special{dvitps: Include0 "psfig.psr"}
\dvitpsLiter@ldim{\drawingwd}%
\dvitpsLiter@ldim{\drawinght}%
\dvitpsLiter@ldim{\psllx bp}%
\dvitpsLiter@ldim{\pslly bp}%
\dvitpsLiter@ldim{\psurx bp}%
\dvitpsLiter@ldim{\psury bp}%
\special{dvitps: Literal "startTexFig"}%
\special{dvitps: Include1 "##1"}%
\special{dvitps: Literal "endTexFig"}%
}}}%
\def\psfordvialw{
\def\PSspeci@l##1##2{
\special{language "PostScript",
position = "bottom left",
literal "  \psllx\space \pslly\space translate
  ##2 1000 div dup scale
  -\psllx\space -\pslly\space translate",
include "##1"}
}}%
\def\psforptips{
\def\PSspeci@l##1##2{{
\d@mx=\psurx bp
\advance \d@mx by -\psllx bp
\divide \d@mx by 1000\multiply\d@mx by \xscale
\incm{\d@mx}
\let\tmpx\dimincm
\d@my=\psury bp
\advance \d@my by -\pslly bp
\divide \d@my by 1000\multiply\d@my by \xscale
\incm{\d@my}
\let\tmpy\dimincm
\d@mx=-\psllx bp
\divide \d@mx by 1000\multiply\d@mx by \xscale
\d@my=-\pslly bp
\divide \d@my by 1000\multiply\d@my by \xscale
\at(\d@mx;\d@my){\special{ps:##1 x=\tmpx cm, y=\tmpy cm}}
}}}%
\def\psonlyboxes{
\def\PSspeci@l##1##2{%
\at(0cm;0cm){\boxit{\vbox to\drawinght
  {\vss\hbox to\drawingwd{\at(0cm;0cm){\hbox{({\tt##1})}}\hss}}}}
}}%
\def\psloc@lerr#1{%
\let\savedPSspeci@l=\PSspeci@l%
\def\PSspeci@l##1##2{%
\at(0cm;0cm){\boxit{\vbox to\drawinght
  {\vss\hbox to\drawingwd{\at(0cm;0cm){\hbox{({\tt##1}) #1}}\hss}}}}
\let\PSspeci@l=\savedPSspeci@l
}}%
%
%
\newread\pst@mpin
\newdimen\drawinght\newdimen\drawingwd
\newdimen\psxoffset\newdimen\psyoffset
\newbox\drawingBox
\newcount\xscale \newcount\yscale \newdimen\pscm\pscm=1cm
\newdimen\d@mx \newdimen\d@my
\newdimen\pswdincr \newdimen\pshtincr
\let\ps@nnotation=\relax
{\catcode`\|=0 |catcode`|\=12 |catcode`|
|catcode`#=12 |catcode`*=14
|xdef|backslashother{\}*
|xdef|percentother{
|xdef|tildeother{~}*
|xdef|sharpother{#}*
}%
\def\R@moveMeaningHeader#1:->{}%
\def\uncatcode#1{%
\edef#1{\expandafter\R@moveMeaningHeader\meaning#1}}%
\def\execute#1{#1}
\def\psm@keother#1{\catcode`#112\relax}
\def\executeinspecs#1{%
\execute{\begingroup\let\do\psm@keother\dospecials\catcode`\^^M=9#1\endgroup}}%
\def\@mpty{}%
\def\matchexpin#1#2{
  \fi%
  \edef\tmpb{{#2}}%
  \expandafter\makem@tchtmp\tmpb%
  \edef\tmpa{#1}\edef\tmpb{#2}%
  \expandafter\expandafter\expandafter\m@tchtmp\expandafter\tmpa\tmpb\endm@tch%
  \if\match%
}%
\def\matchin#1#2{%
  \fi%
  \makem@tchtmp{#2}%
  \m@tchtmp#1#2\endm@tch%
  \if\match%
}%
\def\makem@tchtmp#1{\def\m@tchtmp##1#1##2\endm@tch{%
  \def\tmpa{##1}\def\tmpb{##2}\let\m@tchtmp=\relax%
  \ifx\tmpb\@mpty\def\match{YN}%
  \else\def\match{YY}\fi%
}}%
\def\incm#1{{\psxoffset=1cm\d@my=#1
 \d@mx=\d@my
  \divide\d@mx by \psxoffset
  \xdef\dimincm{\number\d@mx.}
  \advance\d@my by -\number\d@mx cm
  \multiply\d@my by 100
 \d@mx=\d@my
  \divide\d@mx by \psxoffset
  \edef\dimincm{\dimincm\number\d@mx}
  \advance\d@my by -\number\d@mx cm
  \multiply\d@my by 100
 \d@mx=\d@my
  \divide\d@mx by \psxoffset
  \xdef\dimincm{\dimincm\number\d@mx}
}}%
%
\newif\ifNotB@undingBox
\newhelp\PShelp{Proceed: you'll have a 5cm square blank box instead of
your graphics.}%
\def\s@tsize#1 #2 #3 #4\@ndsize{
  \def\psllx{#1}\def\pslly{#2}%
  \def\psurx{#3}\def\psury{#4}
  \ifx\psurx\@mpty\NotB@undingBoxtrue
  \else
    \drawinght=#4bp\advance\drawinght by-#2bp
    \drawingwd=#3bp\advance\drawingwd by-#1bp
  \fi
  }%
\def\sc@nBBline#1:#2\@ndBBline{\edef\p@rameter{#1}\edef\v@lue{#2}}%
\def\g@bblefirstblank#1#2:{\ifx#1 \else#1\fi#2}%
{\catcode`\%=12
\xdef\B@undingBox{
\def\ReadPSize#1{
 \readfilename#1\relax
 \let\PSfilename=\lastreadfilename
 \openin\pst@mpin=#1\relax
 \ifeof\pst@mpin \errhelp=\PShelp
   \errmessage{I haven't found your postscript file (\PSfilename)}%
   \psloc@lerr{was not found}%
   \s@tsize 0 0 142 142\@ndsize
   \closein\pst@mpin
 \else
   \if\matchexpin{\GlobalInputList}{, \lastreadfilename}%
   \else\xdef\GlobalInputList{\GlobalInputList, \lastreadfilename}%
     \immediate\write\psbj@inaux{\lastreadfilename,}%
   \fi%
   \loop
     \executeinspecs{\catcode`\ =10\global\read\pst@mpin to\n@xtline}%
     \ifeof\pst@mpin
       \errhelp=\PShelp
       \errmessage{(\PSfilename) is not an Encapsulated PostScript File:
           I could not find any \B@undingBox: line.}%
       \edef\v@lue{0 0 142 142:}%
       \psloc@lerr{is not an EPSFile}%
       \NotB@undingBoxfalse
     \else
       \expandafter\sc@nBBline\n@xtline:\@ndBBline
       \ifx\p@rameter\B@undingBox\NotB@undingBoxfalse
         \edef\t@mp{%
           \expandafter\g@bblefirstblank\v@lue\space\space\space}%
         \expandafter\s@tsize\t@mp\@ndsize
       \else\NotB@undingBoxtrue
       \fi
     \fi
   \ifNotB@undingBox\repeat
   \closein\pst@mpin
 \fi
\message{#1}%
}%
%
%
\def\psboxto(#1;#2)#3{\vbox{%
   \ReadPSize{#3}%
   \advance\pswdincr by \drawingwd
   \advance\pshtincr by \drawinght
   \divide\pswdincr by 1000
   \divide\pshtincr by 1000
   \d@mx=#1
   \ifdim\d@mx=0pt\xscale=1000
         \else \xscale=\d@mx \divide \xscale by \pswdincr\fi
   \d@my=#2
   \ifdim\d@my=0pt\yscale=1000
         \else \yscale=\d@my \divide \yscale by \pshtincr\fi
   \ifnum\yscale=1000
         \else\ifnum\xscale=1000\xscale=\yscale
                    \else\ifnum\yscale<\xscale\xscale=\yscale\fi
              \fi
   \fi
   \divide\drawingwd by1000 \multiply\drawingwd by\xscale
   \divide\drawinght by1000 \multiply\drawinght by\xscale
   \divide\psxoffset by1000 \multiply\psxoffset by\xscale
   \divide\psyoffset by1000 \multiply\psyoffset by\xscale
   \global\divide\pscm by 1000
   \global\multiply\pscm by\xscale
   \multiply\pswdincr by\xscale \multiply\pshtincr by\xscale
   \ifdim\d@mx=0pt\d@mx=\pswdincr\fi
   \ifdim\d@my=0pt\d@my=\pshtincr\fi
   \message{scaled \the\xscale}%
 \hbox to\d@mx{\hss\vbox to\d@my{\vss
   \global\setbox\drawingBox=\hbox to 0pt{\kern\psxoffset\vbox to 0pt{%
      \kern-\psyoffset
      \PSspeci@l{\PSfilename}{\the\xscale}%
      \vss}\hss\ps@nnotation}%
   \global\wd\drawingBox=\the\pswdincr
   \global\ht\drawingBox=\the\pshtincr
   \global\drawingwd=\pswdincr
   \global\drawinght=\pshtincr
   \baselineskip=0pt
   \copy\drawingBox
 \vss}\hss}%
  \global\psxoffset=0pt
  \global\psyoffset=0pt
  \global\pswdincr=0pt
  \global\pshtincr=0pt 
  \global\pscm=1cm 
}}%
%
%
\def\psboxscaled#1#2{\vbox{%
  \ReadPSize{#2}%
  \xscale=#1
  \message{scaled \the\xscale}%
  \divide\pswdincr by 1000 \multiply\pswdincr by \xscale
  \divide\pshtincr by 1000 \multiply\pshtincr by \xscale
  \divide\psxoffset by1000 \multiply\psxoffset by\xscale
  \divide\psyoffset by1000 \multiply\psyoffset by\xscale
  \divide\drawingwd by1000 \multiply\drawingwd by\xscale
  \divide\drawinght by1000 \multiply\drawinght by\xscale
  \global\divide\pscm by 1000
  \global\multiply\pscm by\xscale
  \global\setbox\drawingBox=\hbox to 0pt{\kern\psxoffset\vbox to 0pt{%
     \kern-\psyoffset
     \PSspeci@l{\PSfilename}{\the\xscale}%
     \vss}\hss\ps@nnotation}%
  \advance\pswdincr by \drawingwd
  \advance\pshtincr by \drawinght
  \global\wd\drawingBox=\the\pswdincr
  \global\ht\drawingBox=\the\pshtincr
  \global\drawingwd=\pswdincr
  \global\drawinght=\pshtincr
  \baselineskip=0pt
  \copy\drawingBox
  \global\psxoffset=0pt
  \global\psyoffset=0pt
  \global\pswdincr=0pt
  \global\pshtincr=0pt 
  \global\pscm=1cm
}}%
%
\def\psbox#1{\psboxscaled{1000}{#1}}%
\newif\ifn@teof\n@teoftrue
\newif\ifc@ntrolline
\newif\ifmatch
\newread\j@insplitin
\newwrite\j@insplitout
\newwrite\psbj@inaux
\immediate\openout\psbj@inaux=psbjoin.aux
\immediate\write\psbj@inaux{\string\joinfiles}%
\immediate\write\psbj@inaux{\jobname,}%
%
%
\def\toother#1{\ifcat\relax#1\else\expandafter%
  \toother@ux\meaning#1\endtoother@ux\fi}%
\def\toother@ux#1 #2#3\endtoother@ux{\def\tmp{#3}%
  \ifx\tmp\@mpty\def\tmp{#2}\let\next=\relax%
  \else\def\next{\toother@ux#2#3\endtoother@ux}\fi%
\next}%
%
%
\let\readfilenamehook=\relax
\def\re@d{\expandafter\re@daux}
\def\re@daux{\futurelet\nextchar\stopre@dtest}%
\def\re@dnext{\xdef\lastreadfilename{\lastreadfilename\nextchar}%
  \afterassignment\re@d\let\nextchar}%
\def\stopre@d{\egroup\readfilenamehook}%
\def\stopre@dtest{%
  \ifcat\nextchar\relax\let\nextread\stopre@d
  \else
    \ifcat\nextchar\space\def\nextread{%
      \afterassignment\stopre@d\chardef\nextchar=`}%
    \else\let\nextread=\re@dnext
      \toother\nextchar
      \edef\nextchar{\tmp}%
    \fi
  \fi\nextread}%
\def\readfilename{\bgroup%
  \let\\=\backslashother \let\%=\percentother \let\~=\tildeother
  \let\#=\sharpother \xdef\lastreadfilename{}%
  \re@d}%
%
%
\xdef\GlobalInputList{\jobname}%
\def\psnewinput{%
  \def\readfilenamehook{
    \if\matchexpin{\GlobalInputList}{, \lastreadfilename}%
    \else\xdef\GlobalInputList{\GlobalInputList, \lastreadfilename}%
      \immediate\write\psbj@inaux{\lastreadfilename,}%
    \fi%
    \let\readfilenamehook=\relax%
    \ps@ldinput\lastreadfilename\relax%
  }\readfilename%
}%
\expandafter\ifx\csname @@input\endcsname\relax    
  \immediate\let\ps@ldinput=\input\def\input{\psnewinput}%
\else
  \immediate\let\ps@ldinput=\@@input
  \def\@@input{\psnewinput}%
\fi%
\def\nowarnopenout{%
 \def\warnopenout##1##2{%
   \readfilename##2\relax
   \message{\lastreadfilename}%
   \immediate\openout##1=\lastreadfilename\relax}}%
\def\warnopenout#1#2{%
 \readfilename#2\relax
 \def\t@mp{TrashMe,psbjoin.aux,psbjoint.tex,}\uncatcode\t@mp
 \if\matchexpin{\t@mp}{\lastreadfilename,}%
 \else
   \immediate\openin\pst@mpin=\lastreadfilename\relax
   \ifeof\pst@mpin
     \else
     \edef\tmp{{If the content of this file is precious to you, this
is your last chance to abort (ie press x or e) and rename it before
retexing (\jobname). If you're sure there's no file
(\lastreadfilename) in the directory of (\jobname), then go on: I'm
simply worried because you have another (\lastreadfilename) in some
directory I'm looking in for inputs...}}%
     \errhelp=\tmp
     \errmessage{I may be about to replace your file named \lastreadfilename}%
   \fi
   \immediate\closein\pst@mpin
 \fi
 \message{\lastreadfilename}%
 \immediate\openout#1=\lastreadfilename\relax}%
{\catcode`\%=12\catcode`\*=14
\gdef\splitfile#1{*
 \readfilename#1\relax
 \immediate\openin\j@insplitin=\lastreadfilename\relax
 \ifeof\j@insplitin
   \message{! I couldn't find and split \lastreadfilename!}*
 \else
   \immediate\openout\j@insplitout=TrashMe
   \message{< Splitting \lastreadfilename\space into}*
   \loop
     \ifeof\j@insplitin
       \immediate\closein\j@insplitin\n@teoffalse
     \else
       \n@teoftrue
       \executeinspecs{\global\read\j@insplitin to\spl@tinline\expandafter
         \ch@ckbeginnewfile\spl@tinline
       \ifc@ntrolline
       \else
         \toks0=\expandafter{\spl@tinline}*
         \immediate\write\j@insplitout{\the\toks0}*
       \fi
     \fi
   \ifn@teof\repeat
   \immediate\closeout\j@insplitout
 \fi\message{>}*
}*
\gdef\ch@ckbeginnewfile#1
 \def\t@mp{#1}*
 \ifx\@mpty\t@mp
   \def\t@mp{#3}*
   \ifx\@mpty\t@mp
     \global\c@ntrollinefalse
   \else
     \immediate\closeout\j@insplitout
     \warnopenout\j@insplitout{#2}*
     \global\c@ntrollinetrue
   \fi
 \else
   \global\c@ntrollinefalse
 \fi}*
\gdef\joinfiles#1\into#2{*
 \message{< Joining following files into}*
 \warnopenout\j@insplitout{#2}*
 \message{:}*
 {*
 \edef\w@##1{\immediate\write\j@insplitout{##1}}*
\w@{
\w@{
\w@{
\w@{
\w@{
\w@{
\w@{
\w@{
\w@{
\w@{
\w@{\string\input\space psbox.tex}*
\w@{\string\splitfile{\string\jobname}}*
\w@{\string\let\string\autojoin=\string\relax}*
}*
 \expandafter\tre@tfilelist#1, \endtre@t
 \immediate\closeout\j@insplitout
 \message{>}*
}*
\gdef\tre@tfilelist#1, #2\endtre@t{*
 \readfilename#1\relax
 \ifx\@mpty\lastreadfilename
 \else
   \immediate\openin\j@insplitin=\lastreadfilename\relax
   \ifeof\j@insplitin
     \errmessage{I couldn't find file \lastreadfilename}*
   \else
     \message{\lastreadfilename}*
     \immediate\write\j@insplitout{
     \executeinspecs{\global\read\j@insplitin to\oldj@ininline}*
     \loop
       \ifeof\j@insplitin\immediate\closein\j@insplitin\n@teoffalse
       \else\n@teoftrue
         \executeinspecs{\global\read\j@insplitin to\j@ininline}*
         \toks0=\expandafter{\oldj@ininline}*
         \let\oldj@ininline=\j@ininline
         \immediate\write\j@insplitout{\the\toks0}*
       \fi
     \ifn@teof
     \repeat
   \immediate\closein\j@insplitin
   \fi
   \tre@tfilelist#2, \endtre@t
 \fi}*
}%
\def\autojoin{%
 \immediate\write\psbj@inaux{\string\into{psbjoint.tex}}%
 \immediate\closeout\psbj@inaux
 \expandafter\joinfiles\GlobalInputList\into{psbjoint.tex}%
}%
%
%
%
\def\centinsert#1{\midinsert\line{\hss#1\hss}\endinsert}%
\def\psannotate#1#2{\vbox{%
  \def\ps@nnotation{#2\global\let\ps@nnotation=\relax}#1}}%
\def\pscaption#1#2{\vbox{%
   \setbox\drawingBox=#1
   \copy\drawingBox
   \vskip\baselineskip
   \vbox{\hsize=\wd\drawingBox\setbox0=\hbox{#2}%
     \ifdim\wd0>\hsize
       \noindent\unhbox0\tolerance=5000
    \else\centerline{\box0}%
    \fi
}}}%
%
\def\at(#1;#2)#3{\setbox0=\hbox{#3}\ht0=0pt\dp0=0pt
  \rlap{\kern#1\vbox to0pt{\kern-#2\box0\vss}}}%
%
\newdimen\gridht \newdimen\gridwd
\def\gridfill(#1;#2){%
  \setbox0=\hbox to 1\pscm
  {\vrule height1\pscm width.4pt\leaders\hrule\hfill}%
  \gridht=#1
  \divide\gridht by \ht0
  \multiply\gridht by \ht0
  \gridwd=#2
  \divide\gridwd by \wd0
  \multiply\gridwd by \wd0
  \advance \gridwd by \wd0
  \vbox to \gridht{\leaders\hbox to\gridwd{\leaders\box0\hfill}\vfill}}%
%
\def\fillinggrid{\at(0cm;0cm){\vbox{%
  \gridfill(\drawinght;\drawingwd)}}}%
%
%
\def\textleftof#1:{%
  \setbox1=#1
  \setbox0=\vbox\bgroup
    \advance\hsize by -\wd1 \advance\hsize by -2em}%
\def\textrightof#1:{%
  \setbox0=#1
  \setbox1=\vbox\bgroup
    \advance\hsize by -\wd0 \advance\hsize by -2em}%
\def\endtext{%
  \egroup
  \hbox to \hsize{\valign{\vfil##\vfil\cr%
\box0\cr%
\noalign{\hss}\box1\cr}}}%
%
\def\frameit#1#2#3{\hbox{\vrule width#1\vbox{%
  \hrule height#1\vskip#2\hbox{\hskip#2\vbox{#3}\hskip#2}%
        \vskip#2\hrule height#1}\vrule width#1}}%
\def\boxit#1{\frameit{0.4pt}{0pt}{#1}}%
\catcode`\@=12 
%
\psfordvips   